\documentclass[a4paper,12pt]{article}

\usepackage{natbib}

\usepackage{sectsty}

\allsectionsfont{\normalsize\bfseries}
\allsectionsfont{\centering}

\usepackage{graphicx}
\usepackage{multirow}
\usepackage{hhline}
\usepackage{times}
\usepackage{lipsum}
\usepackage{wrapfig}
\usepackage{titling}
  \usepackage{color}
\let\OLDthebibliography\thebibliography
\renewcommand\thebibliography[1]{
  \OLDthebibliography{#1}
  \setlength{\parskip}{0pt}
  \setlength{\itemsep}{0pt plus 0.3ex}
}

\begin{document}

{\hskip -0.6cm
\begin{minipage}[h]{16cm}
\textbf{Line Shapes Emitted from Spiral Structures around Symmetric Orbits of Supermassive Binary Black Holes}
\vspace{\baselineskip}

        M.~Smailagi{\' c}$^{a*}$,
        and~E.~Bon$^{a}$
        \vspace{\baselineskip}

$^{a}$ \textit{Astronomical Observatory of Belgrade, Volgina 7, 11060 Belgrade, Serbia}\\

\vspace{\baselineskip}
\end{minipage}

{
\hoffset -3.6cm
\hskip 1.35cm
\begin{minipage}[h]{12cm}
{\textbf{Abstract.} \textit{
Variability of active galactic nuclei is not well understood. One
possible explanation is existence of supermassive binary black
holes (SMBBH) in their centres. 
It is expected that major mergers are common in the Universe.
It is expected that each
supermassive black hole of every galaxy eventually finish as a
SMBBH system in the core of newly formed galaxy. 
Here we model the emission line profiles of active galactic nuclei (AGN) assuming  that the flux and emission line shapes variation are induced 
by supermassive binary black hole systems (SMBBH).
We assume that accreting gas inside of circumbinary (CB) disk is photo ionized by mini accretion disk emission around each SMBBH. 
We calculate variations of emission line flux, shifts and shapes for different parameters of SMBBH orbits. 
We consider cases with different masses and inclinations for circular orbits and measure the effect to the shape of emission line profiles and flux variability.}

}
\end{minipage}
}

\section{Introduction}
\label{sec:1}

One of the mechanisms proposed to explain broad-line region (BLR) 
variability in active galactic nuclei (AGNs) includes existence of SMBBHSs in their centre
\citep[see e.g.][]{Komossa06, Bogdanovic08, Gaskell09, Tsalmantza11, Eracleous12, Popovic12}.
It is believed that nearly all nearby galaxies 
host supermassive black holes (SMBHs) in their centres \citep[][]{Kormendy95, Ferrarese05} 
and that mergers between galaxies are common in the Universe. 
When two galaxies merge, their SMBHs will sink to the centre 
of the newly formed galaxy, due to dynamical friction and interactions with stars and gas,
and, as a consequence, a supermassive binary black hole systems (SMBBHS) with sub-parsec separation will be formed 
\citep[e.g.][]{Begelman80, Colpi11}. 
Depending on how fast the massive galaxies merge, we could expect that a relatively large number of AGNs contain SMBBHSs.

According to standard model type 1 AGN are characterized with prominent broad emission lines (BELs) emitted from the broad line region (BLR) 
\cite[see for example][]{Anton85,Netzer13,Netzer15}. It is a very compact region \cite[see for example][] {Wandel99,Peterson04,Kaspi05,Bentz09,Bentz15}, 
nested in a small dusty torus \cite[see for example][]{Koshida09,Kishimoto11,Stalevski12,Koshida14,Oknyansky14} 
at the core of AGN, where the optical and ultraviolet BELs originate \cite[see for example][]{Sulentic02,Gaskell09,Marz10,Netzer15}. 
Some AGN show single peaked BELs \cite[see][]{Sulentic02,Bon09a,Bon09b,Marz10}
and some show very broad double peaked BELs \cite[see][]{Sulentic02,Strateva03,Bon09b,Marz10}. 
Different models were used to explain double peaked BEL profiles 
\cite[see][]{Gaskell83,ChenHalpern89,Chakrabarti94,Eracleous95,Pringle96,Pop04,Bon06,Bon08,Hayasaki08,Bon09a,Bon09b,Bon10,BonGav10,Lewis10,Shen10,Koll11,Bon12,Eracleous12,Popovic12,Gaskell13,Gaskell14,Goosmann14,Bogdanovic15,Graham15}... 

There exists a number of AGNs with observational signatures of SMBBHSs. 
In some cases, two AGNs are detected via direct imaging in the same galaxy, 
but usually with too wide separations to be gravitationally bound \citep[][]{Komossa03, Comerford13, Liu13, Woo14}
The smallest detected separation is 7 pc \citep[][]{Rodriguez06}.
Indirectly, a number of SMBBHSs are identified from periodic variations in light curves or in light and radial velocity curves 
\citep[e.g.][]{Komossa03, Valtonen08, Boroson09, Bon12, Eracleous12, Ju13, Kun14, Graham15, Komossa15r}. 
SMBBHSs candidates are also identified in other ways, such as 
double-peaked narrow emission lines \citep[][]{Comerford09, Shen11},
the precession of radio jets \citep[][]{Begelman80, Liu07} and  
double - double radio galaxies \citep[][]{Liu03}, 
reverberation mapping \citep[theoretical model,][]{Brem14}.  
However, in many cases alternative explanations also work well, 
and there are currently no confirmed detections of sub-parsec SMBBHSs \citep[][]{Eracleous12}. 
On the other hand, a number of simulations have been performed in order to predicts characteristics of SMBBHSs 
\citep[e.g.][]{Hayasaki08, MacFadyen08, 
     Cuadra09, Shapiro09, Noble12, DOrazio13, Gold13,   
     Farris14, Gold14, Shi15},
 to investigate
the binary torque influences on the surrounding CB disk and what are the radiated spectra. 
In these simulations it is found that between the CB disk and mini-disks surrounding each black hole, a low-density cavities 
and denser spiral forms  of gas are formed
\citep[see e.g.][]{Hayasaki08, MacFadyen08, Cuadra09, Gold13, Shi15}.

     In the previously proposed models
     it was shown that spiral emissivity perturbation is an efficient mechanism of angular momentum transport \citep[e.g.][]{Matsuda89}.   
     However, \citet{Chakrabarti94} first proposed that double-peaked AGNs profile variability could come mainly from the spiral shocks. 
     \citet{Lewis10} used a spiral-arm model \citep[developed by][]{StorchiBergmann03} and showed that the model could explain some (but not all)
     of the characteristics of the observed profiles variability.  
     In addition, there exist a number of simulations that showed that spiral arms form in SMBBHs \citep[e.g.][]{Hayasaki08, MacFadyen08, 
     Cuadra09, Shapiro09, Noble12, DOrazio13, Gold13,   
     Farris14, Gold14, Shi15}.  
     According to these works, accreting gas around a SMBBHS forms
     a cavity inside of CB disk 
     of radius equal to about twice of its semi-major axis, in which accretion streams are found in a form
     two spiral arms. The spiral is stationary in the rotation frame of the SMBBHSs \citep[][]{MacFadyen08}. 
     \citet{DOrazio13} demonstrated that for a mass ratio $Q > 0.05$, which is expected for most SMBBHSs originating from galaxy mergers, 
       the accretion rate is modulated by the SMBBHS, and shows 1-3 distinct periods, while for $Q < 0.05$ the rate shows no variability.
       They also pointed out that for smaller mass ratios (about $Q < 0.3$), the radius of the cavity is smaller.    
    \citet{Shi15} performed a 3D MHD simulation and found that while the cavity forms within a radius of about $2r$, 
    the CB disk extends from $3r$ to $6r$, 
    where $r$ is the semi-major axis. They also found that the accretion rate is the same at a large radius in the disk and onto the binary.

    In this work, we propose a model which could possibly explain observed variability in double-peaked AGNs. 
    Here, we present a simplified model of BELs. The model consists of
supermassive binary black holes on sub parsec orbit,
assuming that on such close distances they would form lower density gaps, with denser spiral features, 
surrounded with CB disk, which would be photoionised by mini-disks around each black hole \cite[see for example][]{Hayasaki08, Gold14}.      
In section 2 we describe the model. In section 3 we present the results, and in 4 we present the conclusions and the discussion.

\section{Model}\label{sec:3}

In our model in the centre of an active galaxy two black holes are orbiting
around their centre of mass and radiation is emitted from mini-disks around each black hole.
We assume that temperatures around mini-disks are constant. 
In the general case, the black holes are orbiting along elliptical orbits which are situated in
one plane and which have a common focus. 
In this work we are considering the case in which orbits have an eccentricity $e=0$, that is, their orbits are circular.
Beyond the spirals a CB disk is situated. 
Between the CB disk and each black hole a spiral of gas is located. The
spirals are rotating with the same angular velocity as the
black holes. 
In our model we assume that continuum
radiation is emitted from mini-disks around black holes, 
while the line forms in spiral arms and CB disk by photoionisation mechanism.
In this model the CB disk extends from $3r$ to $6r$, where $2r$ is the distance between black holes.

\subsection{Description of the model}\label{sec:3.2}

In the Table \ref{tab:1} we summarize parameters and a range of their values for which we have performed the model.
A sketch of the modelled SMBBHS system is presented in the Figure \ref{fig:B}.
In the forthcoming paragraph we describe the model in more details:

\begin{figure}
\begin{center}
\vspace{10mm}
  \includegraphics*[width=8cm]{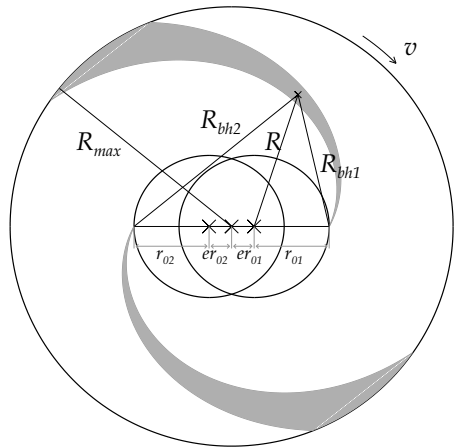} 
\end{center}
\caption{A sketch of a SMBBHS inside the CB disk. Gray areas represent the spiral arms of gas, each of which ends in a SMBH.
Inner circle-like lines present orbits of the SMBHs in 12 time instants. Outer circle presents inner edge of the CB disk. 
Crosses in the middle of the system present centres of the CB disk and of orbits of the SMBHs.
}\label{fig:B}
\end{figure}

\textit{Characteristics of binary black hole system:} Input parameters in our model are
the total mass in the black holes ($M$), their period of rotation ($T$),
the ratio of their masses ($Q=M_{1}/M_{2}$), and the emissivity coefficient of each black  
hole ($q_{1}$ and $q_{2}$). 
It is assumed that mini-disks around each black hole illuminates spiral arms and CB disk, and the emission lines are produced by photoionisation,
, while the emission from cavities are assumed to be negligible.
 The emissivity is equal to 
 
 \begin{equation}
  \epsilon \sim r_{BH1}^{-q_{1}} + r_{BH2}^{-q_{2}} , 
  \label{eq:1}
 \end{equation}
where $r_{BH1,BH2}$ are distances to the black holes. We assume
that $q_{1,2}=2$, which corresponds to the photoionisation. Semi-major axes
of orbits of the black holes ($r_{01}$ and $r_{02}$; $r_{02} \leq
r_{01}$), are calculated from Kepler equations of motion \citep[][]{Hilditch01}:  

 \begin{equation}
  r_{1,2}\sin i = (1.3751 \times 10^{4})(1 - e^{2})^{1/2} K_{1,2} T \hspace{5pt} {\rm
km},
  \label{eq:2} 
 \end{equation}

 \begin{equation}
  M_{1,2}\sin i = (1.0361 \times 10^{-7})(1 - e^{2})^{3/2} (K_{1} +
K_{2})^{2} K_{2,1} T \hspace{5pt} M_{\odot},
  \label{eq:3}
 \end{equation}
where $e$ is the eccentricity of both orbits (for circular orbits it is
$e=0$), $i$ is the angle of inclination, and $K_{1,2}$ are the
semi-amplitudes of the velocity curves.  
In this work we assume that black holes
rotate along circular orbits ($e=0$) with the same radius
$r_{01}=r_{02}=r_{0}$. 
From equations \ref{eq:2} and \ref{eq:3} it could be derived that the radius is equal to  

\begin{equation}
   r_{0}={\rm const}\times M^{1/3}T^{2/3}.  
   \label{eq:4}
\end{equation}

\textit{Definition of spirals and cavities:} 
Spirals are defined
as

 $$R_{1}(\varphi)<R(\varphi + \theta)<R_{2}(\varphi), $$
  $$R \leq k r_{01}, $$
\begin{equation}
 \varphi \geq 0,
 \label{eq:5}
\end{equation}
 where $R$ is the distance from the centre of the orbit of a BH, 
 $\varphi$ is the angle enclosed with x-axis, $k$ is a
parameter which describes the maximum radius to which the spirals
are extending, and $\theta=0$, $\pi$ for the two spirals
inflowing into the black holes. 
Here, for one spiral arm, $R_{1}$ and $R_{2}$ are defined
as 

$$R_{1}=r_{01}e^{b\varphi}$$
\begin{equation}
 R_{2}=r_{02}e^{B\varphi},
 \label{eq:6}
\end{equation}
$b<B$, where $b$ and $B$ are parameters, and $r_{0}$ is the semi-major axis
of the orbit of a black hole (i.e. $r_{0}=r_{01}$ or $r_{0}=r_{02}$).
As $b$ becomes larger spirals become less wrapped. For a fixed
$b$, as $B$ becomes larger the spirals become thicker. 
The other spiral arm is defined in the same way, but with the phase shift of
$\pi$. Parameter $k$ is defined such that the spiral arms are not extended further than $R_{\rm max}$ 
from the common focus, where $R_{\rm max}$ is determined from: 

\begin{equation}
 R_{\rm max}=(a_{01}+a_{02})(1+e)k/2.
 \label{eq:7}
\end{equation}
A CB disk is located between $R_{\rm max}$ and $k_{CB} R_{\rm max}$, where $k_{CB}$ is a parameter.

\textit{Velocities (dynamics of spirals and black holes):}
Velocities of the black holes are calculated from the Kepler's third law, 
with an approximation that the total mass in the black holes is situated 
in the centre of mass of the system: 

\begin{equation}
 v= \bigg( \frac{GM}{R(1-e^{2})} \bigg)^{0.5}\sin i (\cos(\varphi+\omega)+e\cos\omega)
 \label{eq:8}
\end{equation}

Each spiral is rotating as a rigid body with the associated black
hole, but locally their velocities are Keplerian velocities
calculated from the Kepler's third law in the same way.

In addition, local turbulences are taken into account such that
the spectra are smoothed with a Gaussian kernel with width $s$, where $s$ is a parameter. The
larger the parameter $s$ is, the turbulences are stronger and the
spectra are smoother. 

The width of a velocity bin in our model is $400$ ${\rm km/s}$.  

\textit{Angle of view (inclination):} The spectra are viewed
through an angle of inclination $i$, which is an angle between the
plane of rotation of the black holes and the plane normal to the
line of sight. The observed velocities, the smoothing parameter $s$, and
the path which the radiation crosses 
are multiplied by $\sin i$.

\textit{The grid and the resolution in the space and in the time:}
We defined a grid of cells, where each cell is situated between the
lines defined with $R=i_{R}$ and with $b=i_{b}$, where:

 1) The $i_{R}$ is the distance from the centre of mass.
 For $r_{0}\in {r_{01},r_{02}}$ it takes values between $r_{0}$ and $R_{\rm max,i}$,
 such that 
 
 \begin{equation}
  \frac{\ln i_{R} - \ln r_{0}}{\ln R - \ln r_{0}} \times n_{r} = 0,1,2,...,n_{r}-1.
  \label{eq:9}
 \end{equation}
 
 The $n_{r}$ is the resolution in the radial coordinate;

 2) The $i_{b}$ is a parameter which defines division of the spiral arms into smaller spirals. 
 Each spiral arm defined with parameters $b$ and $B$ (see the beginning of this section)
 is divided into a number of 
 smaller (thinner) spirals with parameters $b_{\rm cell}=b_{0}$, $b_{1}$,...,$b_{n_{b}-1}$ and
$B_{\rm cell}=b_{1}$, $b_{2}$,...,$b_{n_{b}}$, where $b_{0}=b$ and
$b_{n_{b}}=B$. We chose equidistant $b_{i}$ parameters. The
$n_{b}$ is the resolution in the $i_{b}$ coordinate, i.e. the number of smaller spirals.

We assume that most of the emission in the line profiles originates from parts of spirals near the black holes. 
Therefore, we use $i_{R}$ which are equidistant in a logarithmic scale, to improve resolution of integration.
Near the black holes the area inside the spiral arms decreases and its emissivity increases
very fast. In every cell flux is calculated by multiplying cell
area with an emissivity from the cell centre. 

The resolution of the grid is $n_{r}\approx 200$ and $n_{b}=90$. The
resolution in the time is 12 (i.e. the number of time instants during
one orbital period for which the spectra are calculated).

As it is believed that nearby galaxy nuclei contain SMBHs with masses $10^{6} - 10^{9} M_{\odot}$ 
\citep[][]{Kormendy95, Ferrarese05}, 
we chose the referred range for the total mass in the SMBBHS system. 

We assume that the CB disk extends from $3r$ to $6r$, 
where $r$ is the semi-major axis, as proposed in \citet{Shi15}. 
This corresponds to $k\sim 3$ and $k_{CB} \sim 2$.

For local turbulences we chose $s=1500 {\rm km/s}$ \citep[see][and references therein]{Lewis10}.

\begin{table*}
\begin{minipage}{150mm}
\centering
\caption{Parameters and their values.} \label{symbols}
\vspace{12pt}
\begin{tabular}{@{}lccc}
\hline Parameter label & Description & A range of values \\
\hline $M$    & total mass in BHs    & $10^{6}-10^{9} M_{\odot}$ \\
       $i$    & inclination          & $45^{\circ}$ ($10^{\circ}-80^{\circ}$)\\
       $T$    & period of rotation   & $\sim 15$ yr\\   
       $s$    & local turbulence     & 1500 ${\rm km/s}$ \\
       $b$    & wrapping of spirals   &  0.45 ($0.1-0.8$)\\ 
       $B-b$  & thickness of spirals  & 0.1 ($0.01-0.35$) \\  
       $q_{1,2}$ & emissivity coefficients & 2 \\
       $k$    & length of spirals ($R_{\rm max}/r_{0}$)   & $3$\\
       $k_{CB}$ & extension of CB disk & 2 \\
\hline
\end{tabular}
\medskip
\label{tab:1}
\end{minipage}
\end{table*}

\section{Results}\label{sec:4}

Here we present calculated spectral line profiles and their characteristics for different parameters.
In Figures \ref{fig:2} - \ref{fig:5} we present how the profiles change with time, and what are the average and the RMS profiles. 
The calculated characteristics of the line profiles include the emitted flux (total, blue, red), the full width at half maximum (FWHM), the centroid velocity, 
and difference between the red and blue flux as a function of time. 
In Figures \ref{fig:6} - \ref{fig:7} we compare how the calculated average values (here defined as a sum of the minimum and the maximum value divided by 2) and amplitudes 
(defined as a difference between the minimum and the maximum value divided by 2) in some of these characteristics change with parameters such as a mass and an inclination. 

\subsection{Spectral line profiles for different masses}\label{sec:4.1}

In Figures \ref{fig:2} - \ref{fig:5} we present 
how the spectra would change when we vary the total mass in the black holes. 
We display spectra for masses $M=10^{6}, 10^{7}, 10^{8}, 10^{9} M_{\odot}$, respectively. 
Other parameters used are $(i, T, b, B, k, s)=(45^{\circ}, 15{\rm yr}, 0.45, 0.55, 2.5, 3.75)$.

For masses ($M \sim 10^{6} M_{\odot}$) the line profiles display a single peak and almost do not change with time. 
  During one orbital period, the FWHM has two maximums and it changes by about 10\%. 
For $M \sim 10^{7} M_{\odot}$ the lines are broader and single-peaked.
As the mass increases, the line profiles become broader, more irregular, with 1-4 peaks, and more variable. 
For $M \sim 10^{8} M_{\odot}$, 
double peak profiles are common, while close to conjunction phase they cannot be resolved,and they appear as a single peak profile.
For masses $M \sim 10^{9} M_{\odot}$ the maximum velocity of the peak is $v \sim 10^{4} {\rm km/s}$, and the line profile 
 sometimes shows multiple peaks.  
For $M \sim 10^{7} - 10^{9} M_{\odot}$ the FWHM has two maximums and it changes 1.5-2 times.

The total, red and blue flux and the centroid velocity are constant in all of these cases. This is expected from the formulation of the model,
as the emitting area does not change (relatively to the SMBHs) and as the system is symmetric. 

As the mass increases, the separation between the black holes also increase, as could be seen from the equation \ref{eq:4}.

\begin{figure*}[h!]
\centering 
  \includegraphics[height=0.4600\textwidth]{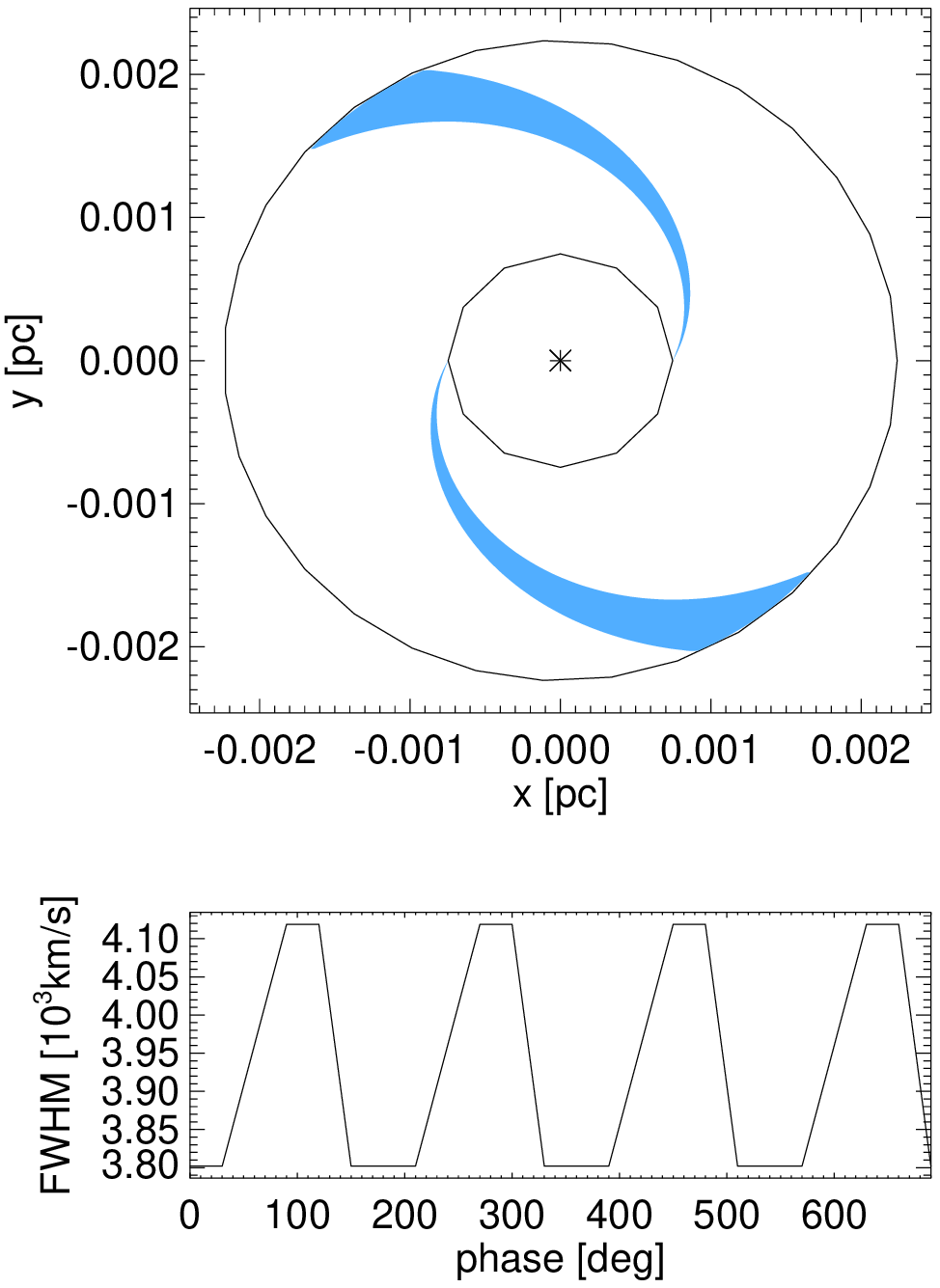}    
   \includegraphics[height=0.4600\textwidth]{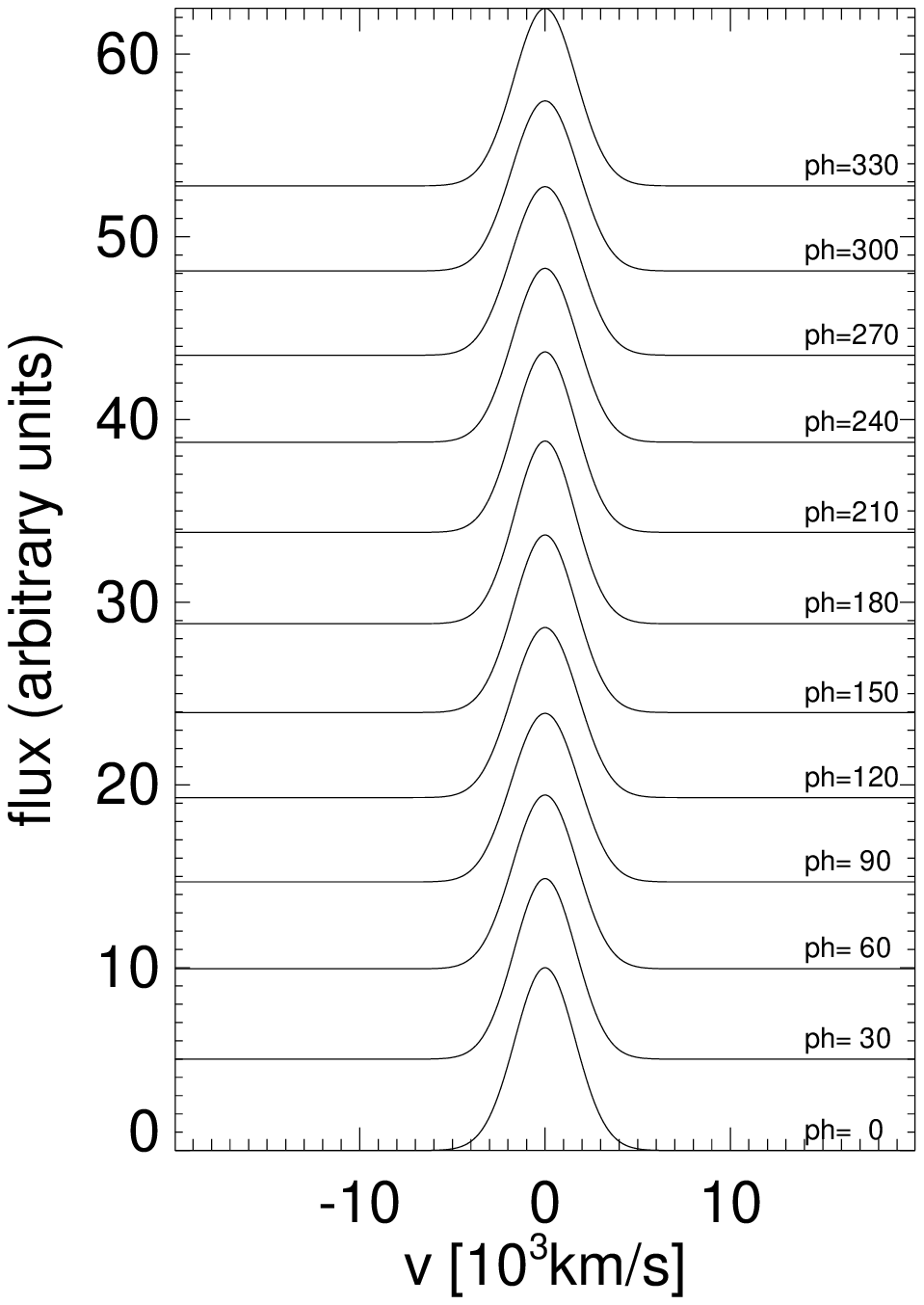}
\caption{\textbf{Upper left:} A sketch of a SMBBHS with the binary black
holes and the spiral arms. 
\textbf{Lower left:} FWHMs as a function of time, during two orbital periods.
\textbf{Right:} Line profiles in a few different time instants during one orbital period. 
The mass in the SMBBHS is $10^{6} M_{\odot}$, and the other parameters used are $(i, T, b,
B, k, s)=(45^{\circ}, 15{\rm yr}, 0.45, 0.55, 2.5, 3.75)$.}
\label{fig:2}
\end{figure*}

\begin{figure*}
\vspace{10mm}
\begin{center}
  \includegraphics*[height=0.4600\textwidth]{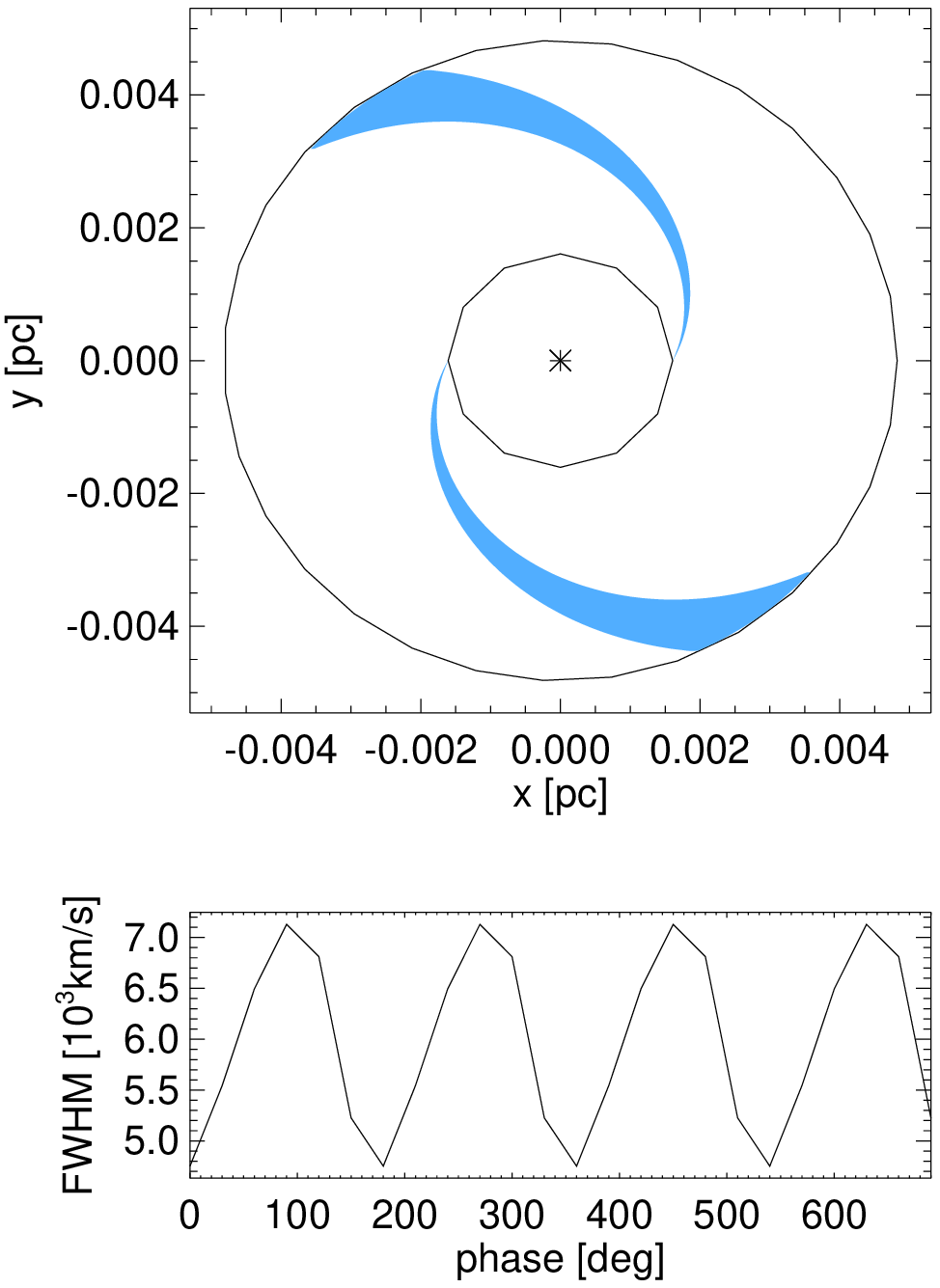}
  \includegraphics*[height=0.4600\textwidth]{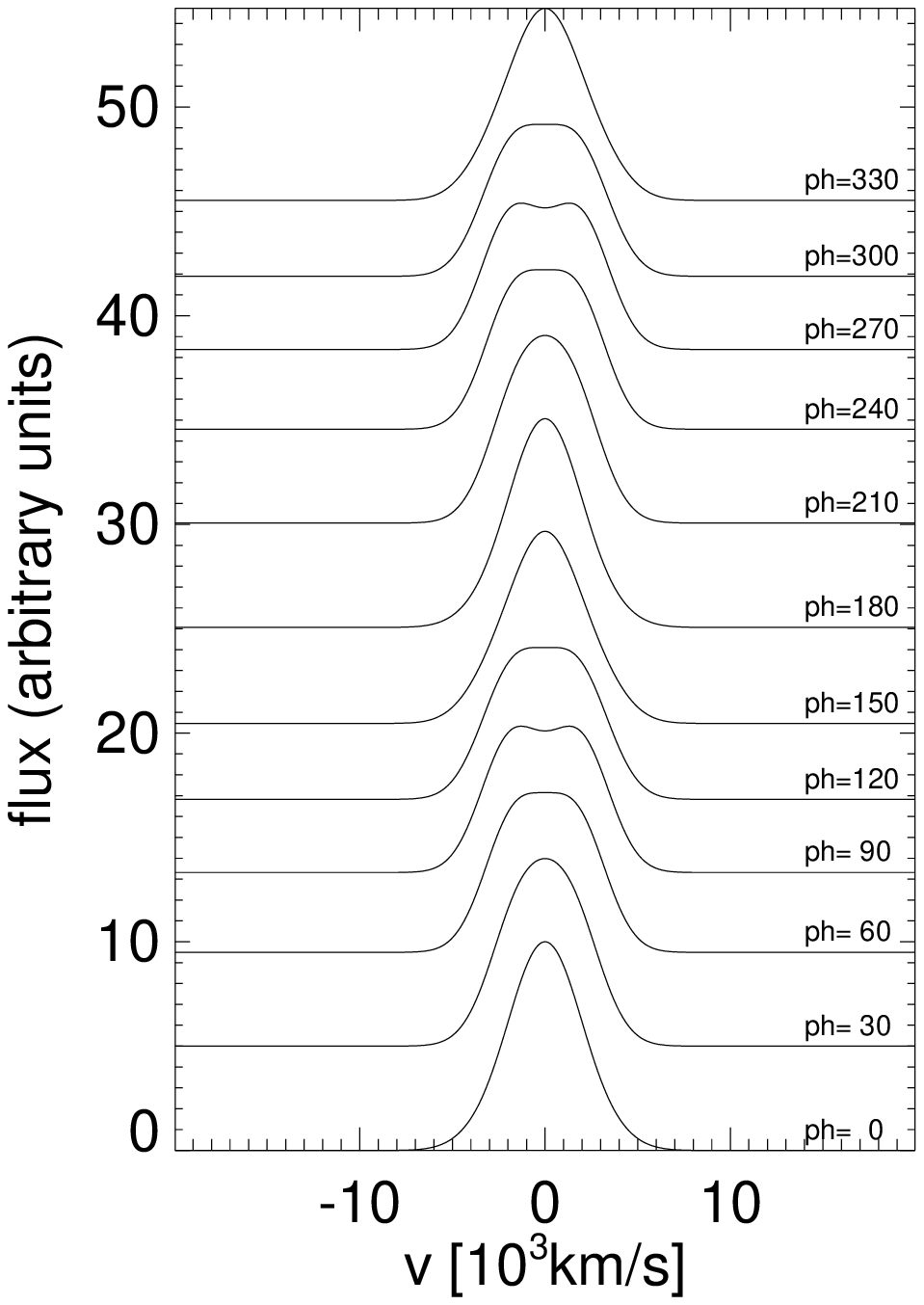}
\end{center}
\caption{Same as in the Figure \ref{fig:2}, but for a SMBBHS with mass of $M = 10^{7} M_{\odot}$.}\label{fig:3}
\end{figure*}

\begin{figure*}
\vspace{10mm}
\begin{center}
  \includegraphics*[height=0.4600\textwidth]{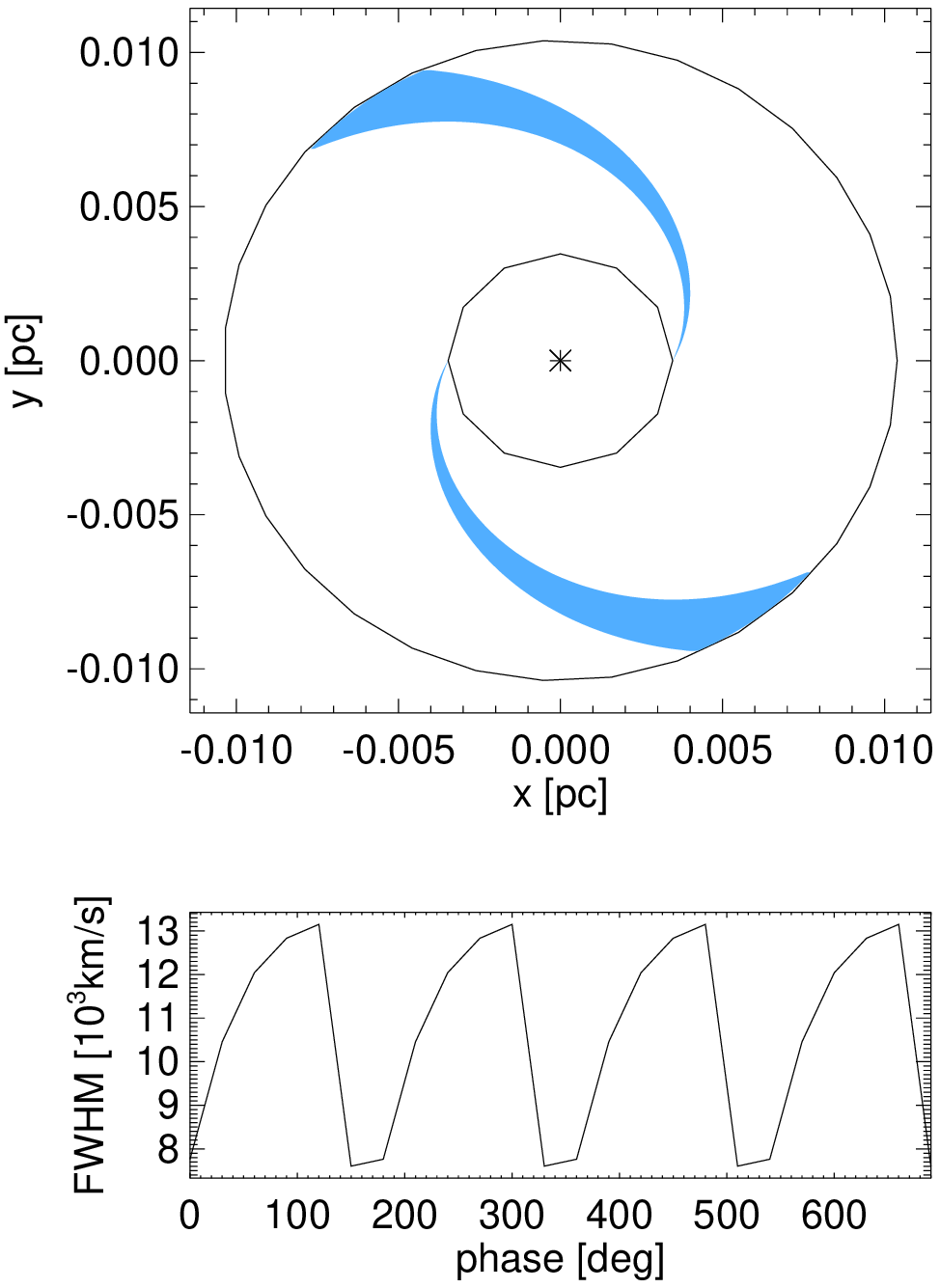}
  \includegraphics*[height=0.4600\textwidth]{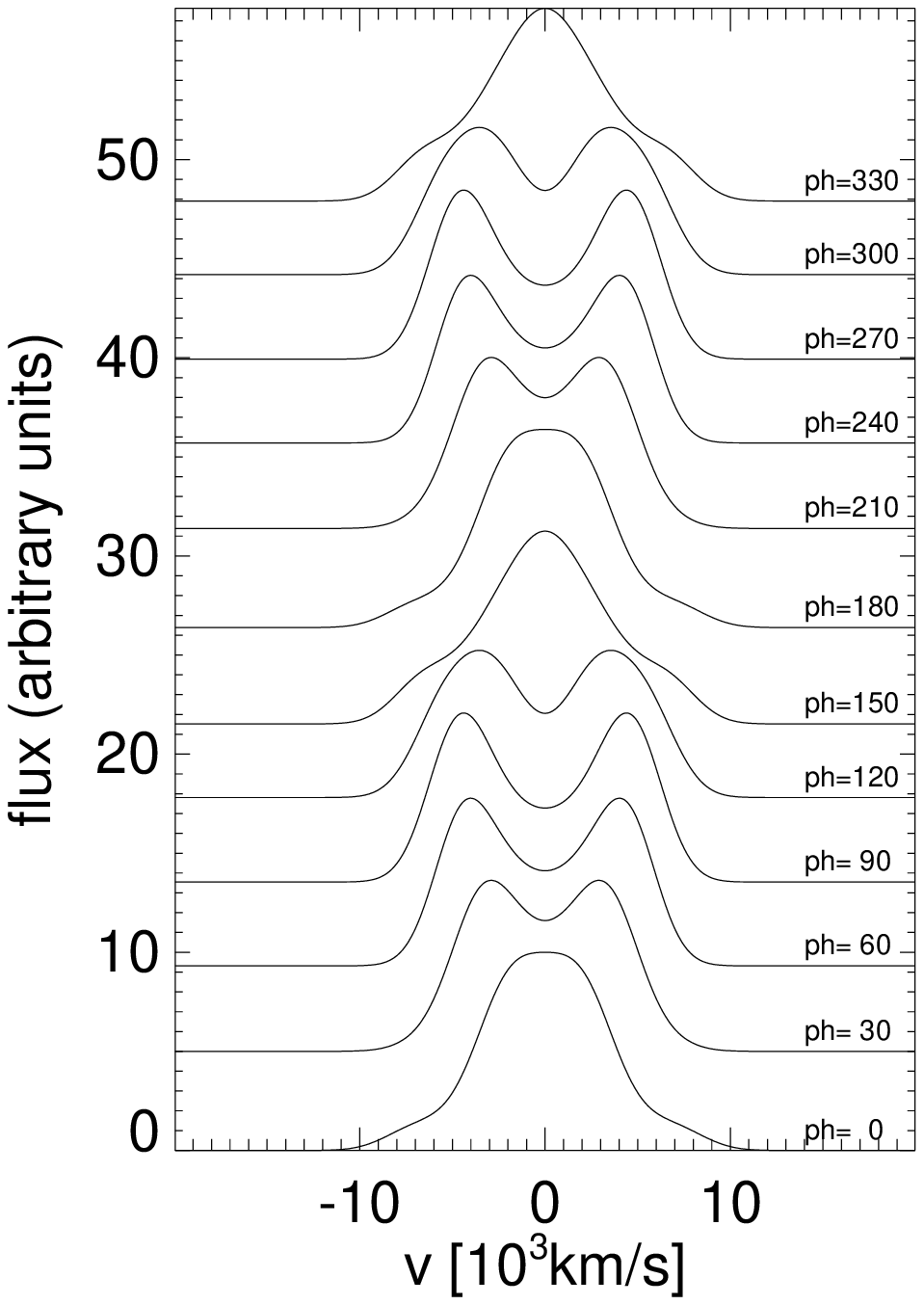}
\end{center}
\caption{Same as in the Figure \ref{fig:2}, but for a SMBBHS with mass of $M = 10^{8} M_{\odot}$.}\label{fig:4}
\end{figure*}

\begin{figure*}
\vspace{10mm}
\begin{center}
  \includegraphics*[height=0.4600\textwidth]{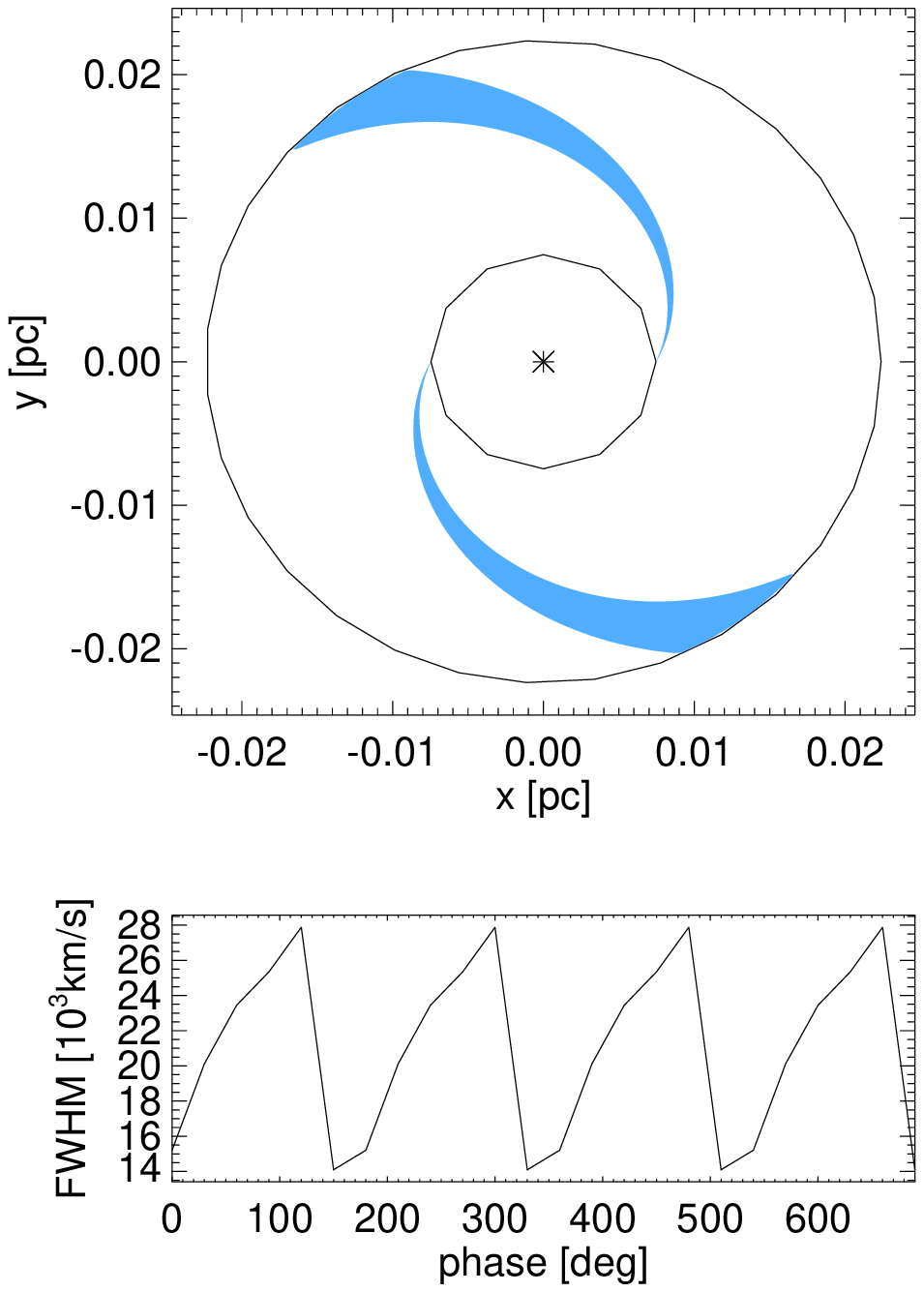}
  \includegraphics*[height=0.4600\textwidth]{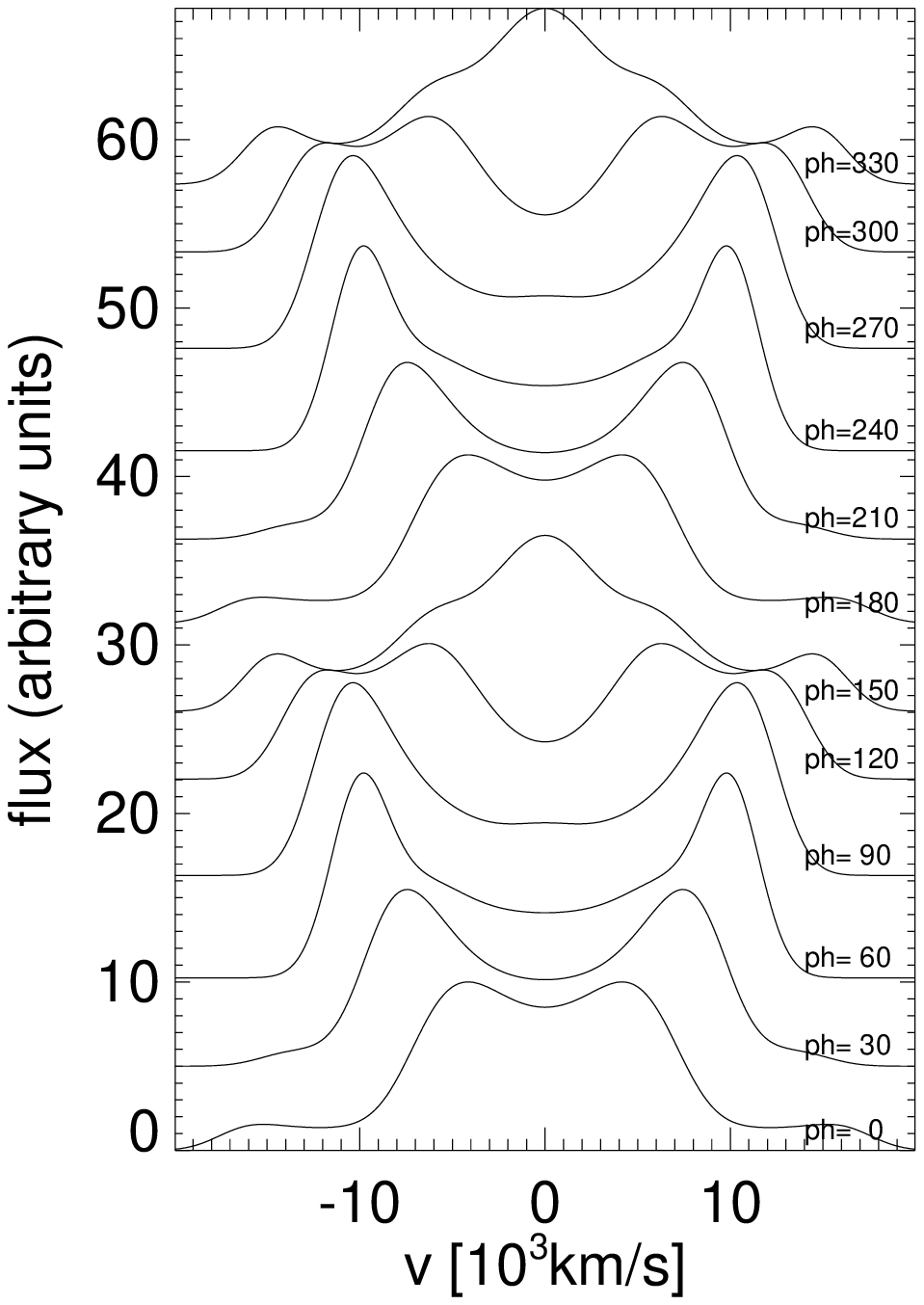}
\end{center}
\caption{Same as in the Figure \ref{fig:2}, but for a SMBBHS with mass of $M = 10^{9} M_{\odot}$.}\label{fig:5}
\end{figure*}

\subsection{FWHM as a function of parameters}\label{sec:4.2}

From equations \ref{eq:8} and \ref{eq:4} it follows that for every cell inside the spirals  
the observed velocity is 
\begin{equation}
   v_{\rm part}\sim {\rm const} \times M^{1/3}T^{-1/3}\sin i,
   \label{eq:10}
\end{equation}

and the observed emissivity is

\begin{equation}
   \epsilon \sim {\rm const} \times M^{-2/3}T^{-4/3}.
   \label{eq:11}
\end{equation}

This implies that, if we ignore local turbulences, a FWHM (and its average value and amplitude) depends on $i$, $T$ and $M$ as 

\begin{equation}
   {\rm FWHM}\sim {\rm const} \times M^{1/3}T^{-1/3}\sin i. 
   \label{eq:12}
\end{equation}

The same relation holds for the centroid velocity $v_{\rm cent}$.
The $v_{\rm cent}$ is defined as follows: we determine half of the maximum emitted flux per wavelength in a line profile, 
then we found the minimum and maximum velocities at which the emitted flux per wavelength is equal to the previously calculated one.

Similarly, from eq. \ref{eq:1} and \ref{eq:4} we have that in every cell emissivities depend on $T$ and $M$ as 

\begin{equation}
   \epsilon \sim {\rm const} \times M^{-2/3}T^{-4/3}, 
   \label{eq:12b}
\end{equation}

and that relative fluxes in the cells are constant with $M$, $i$ and $T$.

In the Figure \ref{fig:6} we present the average value and the amplitude of FWHM as a function of mass, for different inclinations and geometrical parameters. 
As the mass and the inclination increase, the average FWHM and the amplitude in the FWHM would also increase. 
When we calculate FWHMs for different geometrical parameters of the spirals, $0.1<b<0.8$ and $0.05<B-b<0.35$, 
the average FWHMs would change by $\sim \pm 15 \%$.

\begin{figure*}
\begin{center}
\vspace{10mm}
  \includegraphics*[width=8cm]{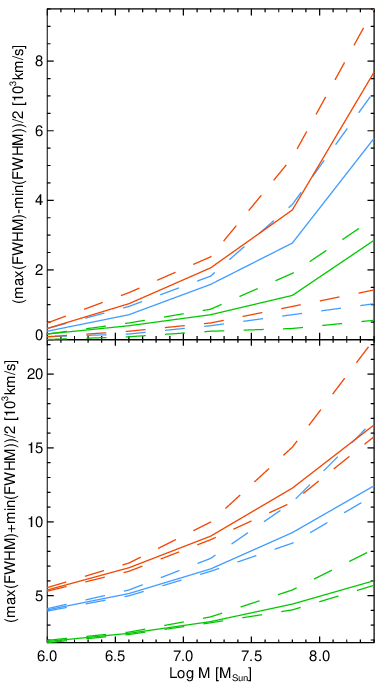} 
\end{center}
\caption{\textbf{Above:} the amplitude in the FWHM as a function of mass. \textbf{Below:} the average value of the FWHM as a function of mass. 
The FWHMs are determined for inclinations of $20^{\circ}$ (green lines), $45^{\circ}$ (blue lines), $70^{\circ}$ (red lines), 
and for a range of geometrical parameters of spirals.
Full lines correspond to $(b,B)=(0.45,0.55)$, while lower (upper) dashed lines are minimums (maximums) in the average values and in the amplitudes in the FWHM, 
for a given mass and inclination, calculated for different geometrical parameters, $0.1<b<0.8$ and $0.05<B-b<0.35$.
}\label{fig:6}
\end{figure*}

\section{Discussion and conclusions}\label{sec:5}

In previous work it was shown that a spiral-arm model could explain some of the characteristics of the observed line profiles variability in some of the AGNs
\citep[see e.g.][]{Lewis10}. 
Motivated by these results, 
in this work we develop a model of a SMBBHS with spiral arms of gas and calculate 
its spectral line profile variations with time, for different parameters, such as mass and inclination. 
By using this model, from the observed variability of emission line profiles (including its FWHM, centroid, flux ratio) 
it is possible to estimate orbital parameters and dynamical masses of their components. 
For example, Figure \ref{fig:6} showed that as mass and inclination increase, average FWHM and amplitude in FWHM variability also increase. 
However, it is not possible to estimate both mass and inclination at the same time.

In this model we assume that the system consists of SMBHs of equal masses on circular orbits.
We measure fluxes, FWHMs, centroid velocities, and red to blue flux ratio as a function of time for 
different masses, inclinations and configurations of the orbits of the black holes, and derive their average values and amplitudes.

 We find that that for some of the AGNs containing SMBBHSs on sub-parsec scale, we could expect that
less massive and less inclined SMBBHSs will display single-peaked line profiles, while more massive and more inclined SMBBHSs will show line profiles with two or more peaks.
We could also expect that 
as the mass and inclination increase, 
the observed lines become wider and more variable, for different geometries of the spiral arms. 
For example, for inclinations of 45$^{\circ}$, SMBBHSs with masses $> 10^{7} M_{\odot}$ will have line profiles with two peaks and widths larger than $\sim 5-7 \times 10^{3}$ km$/$s, 
while for smaller masses the line profiles will have one peak and smaller widths. 

In our paper 2 we will explore how the profiles would change for a system with SMBHs of different masses and for elliptical orbits.  
In future work we will also compare the calculated results with the results of 
  AGNs monitoring programs, in a way similar to \citet{Lewis10}. 
    These results could provide further evidence about SMBBHSs spectral signatures and variability patterns,
    which could be used to identify AGNs which could contain SMBBHSs, and determine their masses and orbital elements.

In the future work \citep[see more in][hereafter paper 2]{paper2} we will explore how the profiles would change for a system with SMBHs of different masses and for elliptical orbits.  
In the future we will also compare the calculated results with the results of 
  AGNs monitoring programs, in a way similar to \citet{Lewis10}. 
  For instance, it would be interesting to compare our results with observed variability in Arp 102B, since 
  \citet{Luka14} found that the variability could not be explained with disk model.
    These results could provide further evidence about SMBBHSs spectral signatures and variability patterns,
    which could be used to identify AGNs which could contain SMBBHSs, and determine their masses and orbital elements.

\section*{Acknowledgment}

The authors would like to thank an anonymous referee and Luka Popovi{\' c} for useful comments.
This work is a part of the project (176001) "Astrophysical Spectroscopy of Extragalactic Objects" 
and (176003) "Gravitation and the large scale structure of the Universe"
supported by the Ministry of Science and Technological Development of Serbia.

\section{Appendix: Influence of CB disk on the calculated line profiles}

Here we analyse effects of CB disk on emission line profiles.
In Figures \ref{fig:7} and \ref{fig:8} we show calculated line profiles in different time instants during two orbital periods for a SMBBHS with a mass of $10^8 M_{\odot}$, 
but considering only
flux emitted in the spiral arms and in the CB disk, respectively. Other parameters are the same as in the Fig. \ref{fig:4}.
Comparing with Fig. \ref{fig:4}, the line profiles have almost the same shape when the flux from the CB disk is excluded. 
The calculated percent of the contribution of the flux emitted from the CB disk is $\sim 23$\%. 
This percent is dependent on the geometry of the spiral arms, it is smaller for thicker and less wrapped spirals (which corresponds to smaller $b$ and larger $B-b$).
For example, for $(b,B-b)=[(0.1,0.01),(0.1,0.35),(0.8,0.01),(0.8,0.35)]$ it is equal to $[9,1,86,18]$ percent.
  
\begin{figure*}[h!]
\vspace{10mm}
\begin{center}
  \includegraphics*[height=0.4600\textwidth]{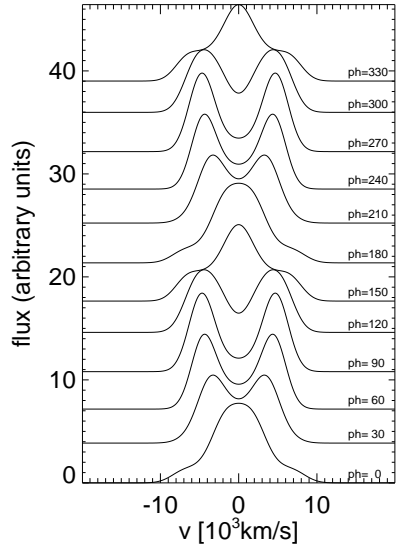}
\end{center}
\caption{Same as in Figure 4, but for flux which originates only from the spiral arms.}\label{fig:7}
\end{figure*}  

\begin{figure*}[h!]
\vspace{10mm}
\begin{center}
  \includegraphics*[height=0.4600\textwidth]{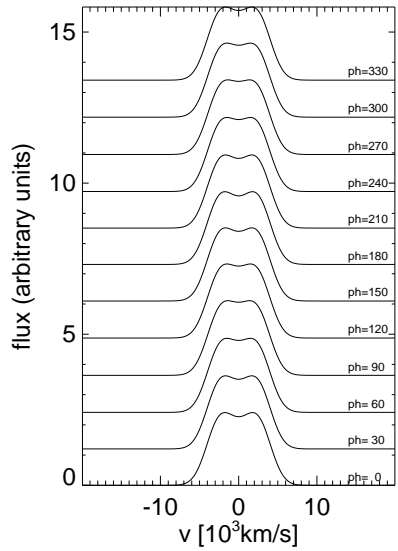}
\end{center}
\caption{Same as in Figure 4, but for flux which originates only from the CB disk.}\label{fig:8}
\end{figure*}

\end{document}